\documentstyle[aps,epsf,12pt]{revtex}

\begin{document}

\begin{titlepage}

\begin{flushright}
\begin{minipage}{4cm}
\begin{flushleft}
SU-4240-652 \\
KEK preprint 96-150 \\
hep-ph/9612358 \\
December, 1996
\end{flushleft}
\end{minipage}
\end{flushright}

\begin{center}
\Large\bf
The Pion Decay Constant and the $\rho $-Meson Mass\\
at Finite Temperature in the Hidden Local Symmetry
\end{center}

\vfil

\begin{center}
{\large
Masayasu {\sc Harada}\footnote{
e-mail address : {\tt mharada@npac.syr.edu}}\\
}
{\it
Department of Physics, Syracuse University, \\
Syracuse, NY 13244-1130, USA \\
}
\ \\
{\large
Akihiro {\sc Shibata}\footnote{%
e-mail address : {\tt ashibata@mail.kek.jp}}\\
}
{\it
National Laboratory for High Enegr Physics (KEK),\\
Tsukuba 305, Japan\\
}
\end{center}

\vfil

\begin{center}
\bf
Abstract
\end{center}

\begin{abstract}
We study the temperature dependence of the pion decay constant and 
$\rho $-meson mass in the hidden local symmetry model at one loop. 
Using the standard imaginary time formalism, we include the thermal
effect of $\rho$ meson as well as that of pion. We show that 
the pion gives a dominant contribution 
to the pion decay constant 
and $\rho $-meson contribution slightly
decreases the critical temperature.
The $\rho$-meson pole mass increases as $T^{4}/m_{\rho }^{2}$ 
at low temperature dominated by the pion-loop effect.
At high temperature, although
the pion-loop effect decreases the $\rho $-meson mass, 
the $\rho $-loop contribution overcomes the pion-loop contribution
and $\rho $-meson mass increases with temperature. We also show that
the conventional parameter
$a$ is stable as the temperature increases.
\end{abstract}

\end{titlepage}

\section{Introduction}

One of the most remarkable feature of QCD in the low energy is that the
approximate chiral symmetry is spontaneously broken, and the approximate
Nambu-Goldstone (NG) boson appears. The pion is regarded as the NG boson and
the resultant low energy theorems successfully reproduce the low energy
physics of pions. In the hot and/or dense matter, however, the quark
condensate melts at some critical point, and the chiral 
symmetry is
restored. This chiral phase transition was widely studied (for recent
reviews, see e.g., Refs.~\cite
{Pisarski:finiteT,Brown-Rho:PRep,Hatsuda-Shiomi-Kuwabara}).

Several experiments such as RHIC are planned to measure the effects in hot
and/or dense matter. One of the interesting quantities in the hot and/or
dense matter is the change of $\rho $-meson mass. In Refs.~\cite
{Brown-Rho:PRep,Brown-Rho:PRL} it was proposed that the $\rho $-meson mass
scaled like the pion decay constant in the hot and/or dense matter, and
vanished at chiral phase transition point. The low temperature theorem
for $\rho $-meson was obtained by using the current algebra\cite
{Dey-Eletsky-Ioffe}, which showed that the $\rho $-meson mass was stable in
the low temperature region. The thermal pion effect to $\rho $-meson mass
was studied by using the effective theories\cite{rhomass:pi}, and it was
shown that the $\rho $ meson mass slightly increased with the temperature.
On the other hand, the thermal effect of nucleon was shown to give a
negative contribution.\cite{Shiomi-Hatsuda}\ 
However, the thermal effects of $%
\rho $-meson itself was not included in these analysis. In Ref.~\cite
{Gerber-Leutwyler:finiteT} the thermal effects of heavier mesons to the
quark condensate were included into the chiral perturbation analysis by
using the dilute gas approximation. The $\rho $ meson seems to give a
non-negligible effect near critical temperature. Then, it is interesting to
see the thermal effect of $\rho $ meson, especially to the $\rho $-meson
mass, by including it systematically.

To include the thermal $\rho $-meson effect systematically it is convenient
to use the effective Lagrangian of the pion and the $\rho $-meson. There are
several models which include pion and $\rho $ meson, among which we study
the hidden local symmetry model.\cite{Bando-Kugo-Yamawaki:PRep} For the
parameter choice $a=2$ this model successfully predicts the following
phenomenological facts\cite{BandoKugoUeharaYamawakiYanagida}: the $\rho $%
-coupling universality, $g_{\rho \pi \pi }=g$\cite{Sakurai}; the KSRF
relation (II), $m_{\rho }^{2}=2g_{\rho \pi \pi }^{2}f_{\pi }^{2}$\cite{KSRF}%
; the $\rho $-meson dominance of the electromagnetic form factor of the
pion, $g_{\gamma \pi \pi }=0$\cite{Sakurai}. Moreover, we obtain the KSRF
(I) relation, $g_{\rho }=2f_{\pi }^{2}g_{\rho \pi \pi }$\cite{KSRF} as a
``low energy theorem'' of hidden local symmetry\cite{BandoKugoYamawaki:NP},
which was first proven at tree level\cite{BandoKugoYamawaki:PTP} and then at
any loop order\cite{Harada-Kugo-Yamawaki}. One loop corrections to the above
predictions were studied in the Landau gauge\cite{Harada-Yamawaki}. The
systematic loop expansion like chiral perturbation was studied in Ref.~\cite
{Tanabashi:hidden:1loop}, where the expansion was done by regarding the $%
\rho $-meson as a light particle.

In this paper we study the temperature dependence of the pion decay constant
and the $\rho$-meson mass by using the hidden local symmetry model at one
loop. 
It is  also
interesting to see the temperature dependence of the parameter $%
a $, which is related to the above successful phenomenological predictions.
One-loop calculation will be done by using the standard imaginary time
formalism.\cite{Matsubara} The renormalization is done in the low energy
limit as shown in Ref.~\cite{Harada-Yamawaki}.

This paper is organized as follows. In section~\ref{sec: HLS}, we briefly
review the hidden local symmetry, and introduce an $R_{\xi }$-like
gauge-fixing term and the corresponding ghost Lagrangian. In section~\ref
{sec: fpi}, we show the temperature dependence of the pion decay constant at
finite temperature. The temperature dependence of the parameter $a$ is also
studied. Section~\ref{sec: mass} is devoted to study the temperature
dependence of the $\rho $-meson mass by using the on-shell--like
renormalization condition. Finally, the summary and discussion are given in
section~\ref{sec: summary}. To avoid complexity, the polarization
tensors at finite temperature and complicated functions are summarized
in Appendices~\ref{app: polar} and \ref{app: functions}, respectively.
We also study the $\rho $-meson propagator at finite temperature 
in Appendix~\ref{app: rho sigma propagator}.

\section{Hidden Local Symmetry}

\label{sec: HLS}

Let us start
with the $[{\rm SU}(N)_{{\rm L}}\times {\rm SU}(N)_{{\rm R}}]_{%
{\rm global}}$$\times $$[{\rm SU}(N)_{{\rm V}}]_{{\rm local}}$ ``linear''
model.\cite{BandoKugoUeharaYamawakiYanagida} We introduce two SU($N$)-matrix
valued variables, $\xi _{{\rm L}}(x)$ and $\xi _{{\rm R}}(x)$, which
transform as 
\begin{equation}
\xi _{{\rm L,R}}(x)\rightarrow \xi _{{\rm L,R}}^{\prime }(x)=h(x)\xi _{{\rm %
L,R}}(x)g_{{\rm R,L}}^{\dag }\ ,
\end{equation}
where $h(x)\in [{\rm SU}(N)_{{\rm V}}]_{{\rm local}}$ and $g_{{\rm L,R}}\in [%
{\rm SU}(N)_{{\rm L,R}}]_{{\rm global}}$. 
These variables are parameterized
as 
\begin{equation}
\xi _{{\rm L,R}}(x)\equiv e^{i\sigma (x)/f_{\sigma }}
e^{\mp i\pi (x)/f_{\pi}}\ ,\quad 
[\pi (x)\equiv \pi ^{a}(x)T_{a}\ ,\ \sigma (x)\equiv
\sigma^{a}(x)T_{a}]\ ,
\end{equation}
where $\pi $ and $\sigma $ are the pion and the ``compensator'' (would-be
Nambu-Goldstone field) to be ``absorbed'' into the hidden gauge boson (the $%
\rho $ meson), respectively, and $f_{\pi }$ and $f_{\sigma }$ are the
corresponding decay constants in the chiral symmetric limit. The covariant
derivatives are defined by 
\begin{eqnarray}
D_{\mu }\xi _{{\rm L}} &\equiv &\partial _{\mu }\xi _{{\rm L}}-igV_{\mu }\xi
_{{\rm L}}+i\xi _{{\rm L}}{\cal L}_{\mu }\ ,  \nonumber \\
D_{\mu }\xi _{{\rm R}} &\equiv &\partial _{\mu }\xi _{{\rm R}}-igV_{\mu }\xi
_{{\rm R}}+i\xi _{{\rm R}}{\cal R}_{\mu }\ ,
\end{eqnarray}
where $g$ is the gauge coupling constant of the hidden local symmetry, $%
V_{\mu }$ ($\equiv V_{\mu }^{a}T_{a}$) the hidden gauge boson field (the $%
\rho $ meson), and ${\cal L}_{\mu }$ and ${\cal R}_{\mu }$ denote the
external fields gauging the $[{\rm SU}(N)_{{\rm L}}]_{{\rm global}}$ and $[%
{\rm SU}(N)_{{\rm R}}]_{{\rm global}}$, respectively. The Lagrangian of $[%
{\rm SU}(N)_{{\rm L}}\times {\rm SU}(N)_{{\rm R}}]_{{\rm global}}$ $\times $ 
$[{\rm SU}(N)_{{\rm V}}]_{{\rm local}}$ ``linear'' model is given by\cite
{Bando-Kugo-Yamawaki:PRep,BandoKugoUeharaYamawakiYanagida}\footnote{%
The massive Yang-Mills approach\cite{Kaymakcalan-Schechter} gives Lagrangian
of the same form
if we take unitary gauge of hidden local symmetry. 
This is equivalent to the hidden local gauge method at
tree level.} 
\begin{equation}
{\cal L}=f_{\pi }^{2}\hbox{tr}\left[ \left( \hat{\alpha}_{\mu \perp }\right)
^{2}\right] +af_{\pi }^{2}\hbox{tr}\left[ \left( \hat{\alpha}_{\mu \parallel
}\right) ^{2}\right] -\frac{1}{2}\hbox{tr}\left[ V_{\mu \nu }V^{\mu \nu
}\right] \ ,  \label{Lag}
\end{equation}
where $a$ is a constant, and $\hat{\alpha}_{\mu \perp }$ and $\hat{\alpha}%
_{\mu \parallel }$ are the covariantized Maurer-Cartan one-forms\cite
{BandoKugoYamawaki:PTP}: 
\begin{eqnarray}
\hat{\alpha}_{\mu \perp } &\equiv &\frac{D_{\mu }\xi _{{\rm L}}\cdot \xi _{%
{\rm L}}^{\dag }-D_{\mu }\xi _{{\rm R}}\cdot \xi _{{\rm R}}^{\dag }}{2i}\ , 
\nonumber \\
\hat{\alpha}_{\mu \parallel } &\equiv &\frac{D_{\mu }\xi _{{\rm L}}\cdot \xi
_{{\rm L}}^{\dag }+D_{\mu }\xi _{{\rm R}}\cdot \xi _{{\rm R}}^{\dag }}{2i}\ .
\end{eqnarray}
Normalizing the kinetic term of $\sigma $, we find\cite{BFY-YINS} 
\begin{equation}
f_{\sigma }^{2}=af_{\pi }^{2}\ .  \label{eq:FsigFpi}
\end{equation}
For the parameter choice $a=2$ at tree level, this model predicts the
following phenomenological facts\cite{BandoKugoUeharaYamawakiYanagida}:

\begin{enumerate}
\item  $g_{\rho \pi \pi }=g$ (universality of the $\rho $-coupling)\cite
{Sakurai};

\item  $m_{\rho }^{2}=2g_{\rho \pi \pi }^{2}f_{\pi }^{2}$ (KSRF II)\cite
{KSRF};

\item  $g_{\gamma \pi \pi }=0$ ($\rho $-meson dominance of the
electromagnetic form factor of the pion)\cite{Sakurai}.
\end{enumerate}

Moreover, independently of the parameter $a$, this model predicts the KSRF
relation\cite{KSRF} (version I) 
\begin{equation}
g_\rho = 2f_\pi^2 g_{\rho\pi\pi}
\end{equation}
as a ``low energy theorem'' of hidden local symmetry\cite
{BandoKugoYamawaki:NP}, which was first proven at tree level\cite
{BandoKugoYamawaki:PTP} and then at any loop order\cite{Harada-Kugo-Yamawaki}%
.

In this paper, to consider the loop effects of hidden local symmetry, we
introduce an $R_\xi$-gauge-like\cite{Fujikawa-Lee-Sanda} gauge-fixing and a
Faddeev-Popov ghost Lagrangian corresponding to the hidden local gauge
boson. These are given by\cite{Harada-Yamawaki} 
\begin{eqnarray}
{\cal L}_{{\rm GF}+{\rm FP}} &=& - \frac{1}{\alpha} \hbox{tr} \left[\left(
\partial^\mu V_\mu \right)\right] + \frac{i}{2} a g f_\pi^2 \hbox{tr}\left[
\partial^\mu V_\mu \left( \xi_{{\rm L}} - \xi_{{\rm L}}^{\dag} + \xi_{{\rm R}%
} - \xi_{{\rm R}}^{\dag} \right) \right]  \nonumber \\
&& {} + \frac{1}{16} \alpha a^2 g^2 f_\pi^4 \left\{ \hbox{tr} \left[ \left(
\xi_{{\rm L}} - \xi_{{\rm L}}^{\dag} + \xi_{{\rm R}} - \xi_{{\rm R}}^{\dag}
\right)^2 \right] - \frac{1}{N} \left( \hbox{tr} \left[ \xi_{{\rm L}} - \xi_{%
{\rm L}}^{\dag} + \xi_{{\rm R}} - \xi_{{\rm R}}^{\dag} \right] \right)^2
\right\}  \nonumber \\
&& {} + i \, \hbox{tr} \left[ \bar{C} \left\{ 2 \partial^\mu D_\mu C + \frac{%
1}{2} \alpha a g^2 f_\pi^2 \left( C \xi_{{\rm L}} + \xi_{{\rm L}}^{\dag} C +
C \xi_{{\rm R}} + \xi_{{\rm R}}^{\dag} C \right) \right\} \right] \ ,
\label{Lag: GF FP}
\end{eqnarray}
where $\alpha$ denotes a gauge parameter and $C$ denotes a ghost field. In
this paper we choose the Landau gauge, $\alpha=0$. In this gauge the
would-be Nambu-Goldstone (NG) boson $\sigma$ is still massless, no other
vector-scalar interactions are created and the ghost field couples only to
the hidden local gauge field.

The renormalization is done by introducing counter terms.\cite
{Harada-Yamawaki,Tanabashi:hidden:1loop} Here following Ref.~\cite
{Harada-Yamawaki}, we use the $Z$ factors defined by 
\begin{equation}
\begin{array}{ll}
g_{0}=Z_{g}g\ , & V_{0\mu }=Z_{V}^{1/2}V_{\mu }\ , \\ 
\pi _{0}=Z_{\pi }^{1/2}\pi \ , & \sigma _{0}=Z_{\sigma }^{1/2}\sigma \ , \\ 
f_{\pi 0}=Z_{\pi }^{1/2}f_{\pi }\ , & f_{\sigma 0}=Z_{\sigma
}^{1/2}f_{\sigma }\ .
\end{array}
\label{Z factors}
\end{equation}
We renormalize the theory at zero temperature, then the corrections from the
thermal pion, $\rho $ meson, and so on are calculated to be finite.\footnote{%
To perform the complete renormalization at one loop, we need higher order
counter terms.\cite{Tanabashi:hidden:1loop} However, for the quantities we
are studying in this paper, the counter terms induced from $Z$-factors in
Eq.~(\ref{Z factors}) are enough to renormalize the divergences. Then we do
not explicitly introduce such higher order counter terms here.}

\section{The Temperature Dependence of Decay Constants}

\label{sec: fpi}In this section we study the temperature dependences of the
decay constants of pion and sigma. We also study the temperature dependence
of the parameter $a$, which is related to the successful phenomenological
predictions of hidden local symmetry.

To define the pion decay constant at finite temperature, we consider a Green
function for the axial-vector current $A_{a}^{\mu }$ 
\begin{equation}
\delta _{ab}G_{{\cal A}}^{\mu \nu }(p_{0},\vec{p};T)\equiv \mbox{\rm F.T.}%
\bigl\langle 0\bigl\vert A_{a}^{\mu }A_{b}^{\nu }\bigr\vert 0\bigr\rangle%
_{T}\ .
\end{equation}
Since the axial-vector current is conserved, we can decompose this Green
function into longitudinal and transverse pieces: 
\begin{equation}
G_{{\cal A}}^{\mu \nu }(p_{0},\vec{p};T)=P_{T}^{\mu \nu }G_{{\cal A}%
T}+P_{L}^{\mu \nu }G_{{\cal A}L}\ ,
\end{equation}
where polarization tensors $P_{L}$ and $P_{T}$ are defined in Eq.~(\ref
{polar tensor}). It is natural to define the pion decay constant at finite
temperature through the longitudinal component in the low energy limit%
\footnote{%
Even when we use the transverse part instead of the longitudinal part to
define $f_\pi(T)$ in Eq.~(\ref{def: fpi}), we obtain the same result: $G_{%
{\cal A}T}(p_0,\vec{p}=0) = G_{{\cal A}L}(p_0,\vec{p}=0)$.}:\cite
{Bochkarev-Kapusta} 
\begin{equation}
f_{\pi }^{2}(T)\equiv -\lim_{p_{0}\rightarrow 0}G_{{\cal A}L}(p_{0},\vec{p}%
=0)\ .  \label{def: fpi}
\end{equation}
There are two types of contributions to this Green function in our model:
(i) the pion exchange diagrams, and 
(ii) the contact or one-particle irreducible (1PI) diagrams. 
The contribution (i) is proportional to 
$p_{\mu }$ or $p_{\nu }$ at one loop. 
At most only one of
the $A_{a}^{\mu }$-$\pi $
coupling can be corrected at one loop, which is not generally proportional
to four momentum $p_{\mu }$. The other coupling is the tree-level one and
proportional to $p_{\mu }$. 
When we act with the projection operator $P_{L\mu \nu}$, 
the term proportional to $p_{\mu }$ vanishes. Due to the current
conservation we have same kinds of contributions from 1PI diagrams: those
are roughly proportional to $g_{\mu \nu }$ instead of $p_{\mu }$. Then we
will calculate only the 1PI diagrams.

\begin{figure}[tbph]
\begin{center}
\ \epsfbox{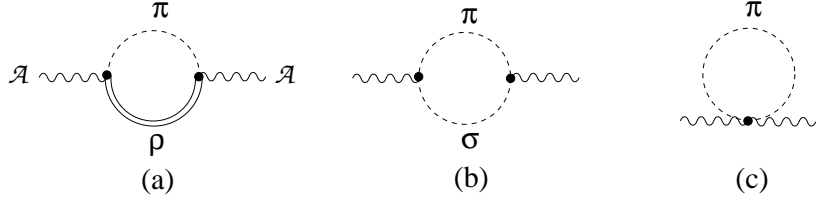}
\end{center}
\caption[]{ Feynman diagrams contributing to the one particle
irreducible part of the axial-vector current two-point function.}
\label{fig: aa}
\end{figure}
There exist three 1PI diagrams which contribute to $G_{{\cal A}}^{\mu \nu }$
at one-loop level: (a) $\pi $+$\rho $ loop, (b) $\pi $+$\sigma $ loop, (c) $%
\pi $ tad-pole, which are shown in Fig.~\ref{fig: aa}. 
These diagrams include ultraviolet divergences, which are renormalized by $%
Z_{\pi }$ in Eq.~(\ref{Z factors}). By taking a suitable subtraction scheme
at zero temperature, all the divergences including finite corrections in the
low energy limit are absorbed into $Z_{\pi }$.\cite
{Harada-Kugo-Yamawaki,Harada-Yamawaki} Then the loop diagrams generate only
the temperature dependent part. By using standard imaginary time formalism%
\cite{Matsubara} we obtain 
\begin{eqnarray}
G_{{\cal A}L}^{(a)}(p_{0},\vec{p} =0)
&=&\frac{N}{2}\frac{a}{\pi ^{2}}\left[ \frac{5%
}{6}I_{2}-J_{1}^{2}+\frac{1}{3m_{\rho }^{2}}\left( I_{4}-J_{1}^{4}\right)
\right] ,  \nonumber \\
G_{{\cal A}L}^{(b)}(p_{0},\vec{p} =0)
&=&\frac{N}{2}\frac{a}{6\pi ^{2}}I_{2}\ ,
\label{eq:fpiT} \\
G_{{\cal A}L}^{(c)}(p_{0},\vec{p} =0)
&=&\frac{N}{2}\frac{1-a}{\pi ^{2}}I_{2}\ , 
\nonumber
\end{eqnarray}
where the functions $I_{n}$ and $J_{m}^{n}$ are defined in appendix B. If we
saw diagrams naively, we might think the diagram 
(c) gives the dominant
contribution. However, each diagram does generate the dominant contribution $%
I_{2}=(\pi^{2}/6)T^{2}$. Nevertheless, we see from Eq. (\ref{eq:fpiT})
that the terms proportional to $aI_{2}$ are cancelled among three diagrams.
The total contribution is given by 
\begin{equation}
f_{\pi }^{2}(T)=f_{\pi }^{2}-\frac{N}{2}\frac{1}{\pi ^{2}}\left[
I_{2}-aJ_{1}^{2}+\frac{a}{3m_{\rho }^{2}}\left( I_{4}-J_{1}^{4}\right)
\right] \; .  \label{eq: fpi}
\end{equation}
When we take $m_{\rho }\rightarrow \infty $ limit in 
the above expression,  only
the $I_{2}$-term remains; 
\begin{equation}
\left[ f_{\pi }^{{\rm ChPT}}(T)\right] ^{2}=f_{\pi }^{2}-\frac{N}{2}\frac{%
I_{2}}{\pi ^{2}}=f_{\pi }^{2}-\frac{N}{12}T^{2}\ ,  \label{fpi: ChPT}
\end{equation}
which is consistent with the result given by Gasser-Leutwyler. \cite
{Gasser-Leutwyler:finiteT:1987}

\begin{figure}[htbp]
\begin{center}
\ \epsfbox{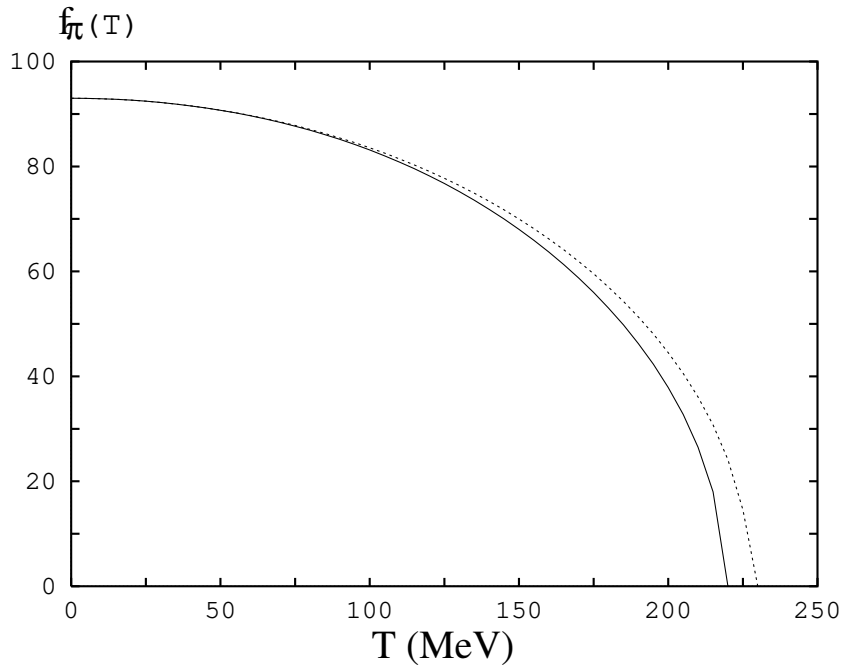}
\caption[]{ The temperature dependence of the pion decay constant. The
solid line denotes the total contribution, $f_{\pi }(T)$ in
Eq.~(\ref{eq: fpi}) 
and the dotted line denotes the $m_{\rho }\rightarrow \infty $ limit 
(Eq.~(\ref{fpi: ChPT})).}
\label{fig: fpi}
\end{center}
\end{figure}

We show the temperature dependence of $f_{\pi }(T)$ by a solid line in Fig.~%
\ref{fig: fpi}, where we use $f_{\pi }=93$\thinspace MeV, $m_{\rho }=770$%
\thinspace MeV, $a=2$ and $N=2$. (We use same values for numerical analysis
below.) 
The chiral perturbation prediction Eq.(\ref{fpi: ChPT}) is shown by a dotted
line in Fig.~\ref{fig: fpi}. Figure~\ref{fig: fpi} shows that the pion loop
gives a dominant contribution and $\rho $-meson loop generates a small
correction. The situation is similar to the quark condensate in the chiral
perturbation analysis.\cite{Gerber-Leutwyler:finiteT} 
We should notice that the
term proportional
to $I_{4}$ corresponds to a part of two-loop order effects in the 
ordinary
chiral perturbation theory. The two-loop order effects are divided into two
types of diagrams: one is two-loop diagram which includes $O(p^{2})$
vertices only and the other is one-loop diagram which includes only one $%
O(p^{4})$ vertex. In chiral perturbation theory the $O(p^{4})$ vertices are
parameterized by $\ell _{1}$ and $\ell _{2}$\cite{Gasser-Leutwyler:SU(2)},
which are saturated by the effect of $\rho $ meson.\cite{rhodominance} Then
the $I_{4}$ term generated by $\rho $ meson may be a good approximation to $%
O(p^{4})$ contribution in the chiral perturbation theory.

Next, we study the temperature dependence of $f_\sigma$. Similar to the case
of $f_\pi$, we start with a Green function for the vector current: 
\begin{equation}
\delta_{ab} G_{{\cal V}}^{\mu\nu}(p_0,\vec{p};T) \equiv \mbox{\rm F.T.}\,
\langle 0 \vert {\cal V}_a^\mu {\cal V}_b^\nu \vert 0 \rangle_T \ .
\end{equation}
There are three types of contributions: (i) 1PI diagrams; 
(ii) the $\sigma$-exchange diagrams; 
(iii) the $\rho$-exchange diagrams. We have no strict
definition of $f_\sigma(T)$ like $f_\pi(T)$ in Eq.~(\ref{def: fpi}), since $%
\sigma$ is not a physical particle. The $\sigma$ is a would-be
Nambu-Goldstone boson which is absorbed into $\rho$ meson. In the Landau
gauge, the $\rho$-exchange contribution is separately conserved,\footnote{%
At tree level, the $\rho$-$\gamma$ mixing is proportional to $g_{\mu\nu}$.
One-loop effects at finite temperature violate this structure, and generally
the $\rho$-$\gamma$ mixing is decomposed into four independent
polarizations: $\Pi_{\rho {\cal V}}^{\mu\nu} = P_T^{\mu\nu} 
\Pi_{\rho {\cal V}T} +
P_L^{\mu\nu} \Pi_{\rho {\cal V}L} + 
P_C^{\mu\nu} \Pi_{\rho {\cal V}C} + 
P_D^{\mu\nu} \Pi_{\rho {\cal V}D}$, 
where the polarization tensors are given in Eq.~(\ref{polar tensor}). 
In the low momentum limit we can easily show that 
$\Pi_{\rho {\cal V}C}$
vanishes. Since $P_{D\mu\alpha} P_L^{\alpha\nu} = P_{D\mu\alpha}
P_T^{\alpha\nu} = P_{L\mu\alpha} P_T^{\alpha\nu} = 0$, $P_{T\mu\alpha}
P_T^{\alpha\nu} = - P_{T\mu}^{\nu}$ and $P_{L\mu\alpha} P_L^{\alpha\nu} = -
P_{L\mu}^{\nu}$, the $\rho$ meson exchange diagrams generate only the terms
proportional to $P_{L\mu\nu}$ and $P_{T\mu\nu}$ to the vector-current
two-point function. These polarization tensors vanish if they are multiplied
by $p^\mu$. This implies that the contribution from the $\rho$-meson
exchange diagrams vanishes if it is multiplied by $p^\mu$, 
and then both the 
$\rho$-meson exchange diagrams and $\sigma$-exchange plus 1PI diagrams
conserve separately.} 
as we can show easily by using the $\rho$ propagator 
(\ref{rho propagator 2}). {}From the conservation of whole current, the sum
of 1PI and $\sigma$-exchange diagrams are conserved. Then, it is reasonable
to define $f_\sigma(T)$ like Eq.~(\ref{def: fpi}) through only the 1PI and $%
\sigma$-exchange diagrams: 
\begin{equation}
f_\sigma^2(T) \equiv - \lim_{p_0\rightarrow0} G_{{\cal V}L}^{({\rm 1PI}%
+\sigma)} (p_0, \vec{p}=0;T) \ .  \label{def: fsigma}
\end{equation}
As discussed below Eq.~(\ref{def: fpi}) for $f_\pi(T)$, it is enough to
calculate the 1PI diagrams to determine the temperature dependence of $%
f_\sigma(T)$.

\begin{figure}[htbp]
\begin{center}
\ \epsfbox{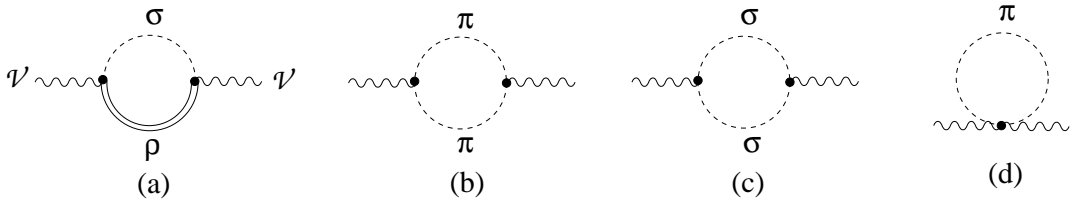}
\end{center}
\caption[]{ Feynman diagrams for the one particle irreducible part of
the vector current two-point function.}
\label{fig: vector current}
\end{figure}
There are four diagrams which contribute to the 1PI part of $G_{{\cal V}%
}^{\mu\nu}$ at one loop, which are shown in Fig.~\ref{fig: vector current}. 
We note that we do not have $\sigma$ tad-pole contribution, since there is
no $\sigma$-$\sigma$-${\cal V}$-${\cal V}$ type vertex at tree level. These
four diagrams lead to the temperature dependence given by 
\begin{equation}
f_\sigma^2(T) = f_\sigma^2 - \frac{N}{2} \frac{1}{\pi^2} \left[ \frac{%
a^2+8a+3}{12} I_2 - J_1^2 + \frac{1}{3m_\rho^2} \left( I_4 - J_1^4 \right)
\right] \ .  \label{fsigma: T}
\end{equation}

\begin{figure}[tbph]
\begin{center}
\ \epsfbox{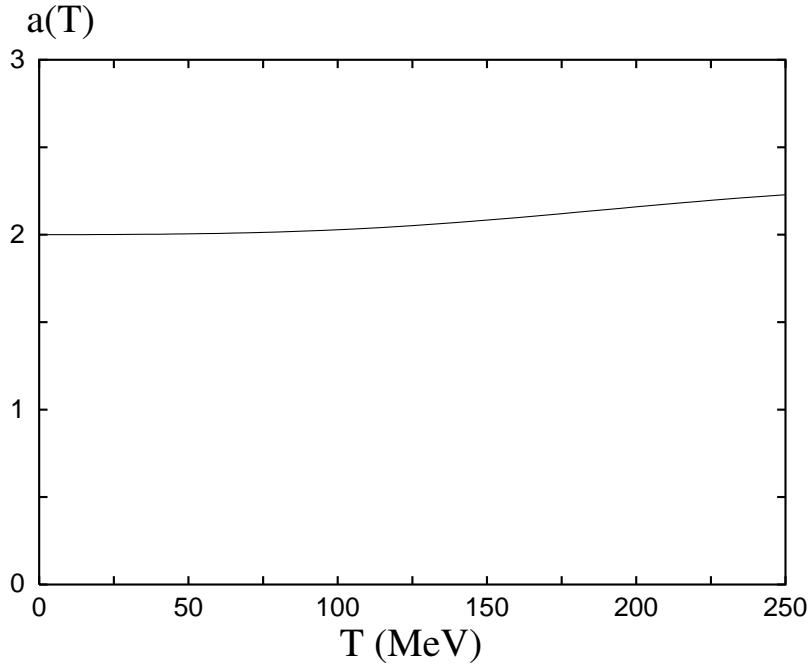}
\end{center}
\caption[]{ The temperature dependence of the parameter $a$.}
\label{fig: apara}
\end{figure}

The parameter $a$ at finite temperature is given by $a(T)=f_{\sigma
}^{2}(T)/f_{\pi }^{2}(T)$ (see Eq. (\ref{eq:FsigFpi})). 
Using the temperature
dependence of $f_{\pi }$ and $f_{\sigma }$ given in Eqs.~(\ref{eq: fpi}) and
(\ref{fsigma: T}), we obtain the temperature dependence of the parameter $a$%
: 
\begin{equation}
\frac{a(T)}{a(0)}=1+\frac{N}{2}\frac{1}{\pi ^{2}f_{\pi }^{2}}\left[ -\frac{%
(a-1)(a-3)}{12a}I_{2}+\frac{1-a^{2}}{a}J_{1}^{2}-\frac{1-a^{2}}{3am_{\rho
}^{2}}\left( I_{4}-J_{1}^{4}\right) \right] \ .  \label{a def: B}
\end{equation}
We show the temperature dependence of the parameter $a$ in 
Fig.~\ref{fig: apara}, 
where we take $a(0)=2$.
The parameter $a$ does not change very much against the temperature, 
i.e., $a(T)\simeq 2$ for wide range of the temperature. 
It was shown\cite
{Harada-Yamawaki} that at zero temperature one-loop effects did not generate
the direct $\gamma \pi \pi $ vertex if and only if we took a parameter
choice $a=2$. 
The fact that $a(T)\simeq 2$ for wide range may suggest that
the successful phenomenological predictions of hidden local symmetry 
discussed in section~\ref{sec: HLS}
hold at finite temperature in good approximation.
We should note that if
we take $a=1$ from the beginning the parameter $a$ does not change with
temperature. Together with the fact that the parameter $a$ is not
renormalized for the parameter choice $a=1$, this implies that in the
``vector limit''\cite{Georgi:vectorlimit} the thermal effects of the $\rho $
meson do not induce deviation from $a=1$.

\section{The Temperature Dependence of $\rho$ Meson Mass}

\label{sec: mass}In this section we study the temperature dependence of $%
\rho $-meson mass. As we show in Appendix~\ref{app: rho sigma propagator},
the $\rho $ and $\sigma $ propagators are separated with each other in the
Landau gauge, and the $\rho $ propagator takes easy form:\footnote{%
Shiomi and Hatsuda \cite{Shiomi-Hatsuda}  used
similar form, where they
started from the Steukelberg formalism.} 
\begin{equation}
D_{\mu \nu }=-\frac{P_{T\mu \nu }}{p^{2}-m_{\rho }^{2}+\Pi _{T}}-\frac{%
P_{L\mu \nu }}{p^{2}-m_{\rho }^{2}+\Pi _{L}}\ .  \label{rho propagator}
\end{equation}
It is reasonable to define the $\rho $-meson mass by using the longitudinal
part.\footnote{%
It should be noticed that the transverse polarization agrees with the
longitudinal one in the low momentum limit: $\Pi_{\rho T}(p_0,\vec{p}=0;T) =
\Pi_{\rho L}(p_0,\vec{p}=0;T)$.}

\begin{figure}[htbp]
\begin{center}
\ \epsfbox{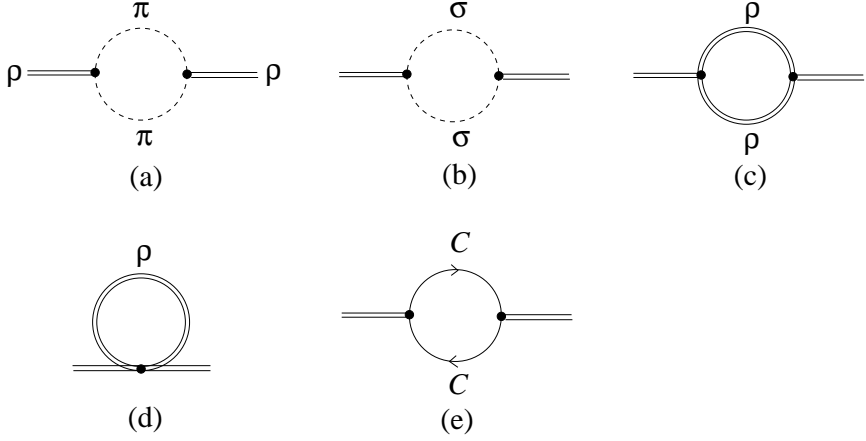}
\end{center}
\caption[]{ Feynman diagrams contributing to the 
$\rho$-meson self-energy: a) $\pi$ loop, 
b) $\sigma$ loop, c) $\rho$ loop, d) $\rho$ tad-pole and e) ghost
loop. }
\label{fig: rho propagator}
\end{figure}
The one-loop diagrams contributing $\rho$ meson self-energy are shown in
Fig.~\ref{fig: rho propagator}. 
In Ref.~\cite{Harada-Yamawaki}, the finite term of $Z_VZ_g^2$ was determined
but each finite term of $Z_V$ and $Z_g$ was not determined separately. We
introduce the parameter $z$ for expressing this finite term as 
\begin{eqnarray}
Z_{V}-1 &=&\frac{g^{2}}{(4\pi )^{2}}\left[ \frac{51-a^{2}}{12}\left\{ \frac{1%
}{\overline{\epsilon }}-\ln m_{\rho }^{2}+\frac{5}{6}\right\} +z\right] \ , 
\nonumber \\
Z_{g}-1 &=&\frac{g^{2}}{(4\pi )^{2}}\left[ -\frac{87-a^{2}}{24}\left\{ \frac{%
1}{\overline{\epsilon }}-\ln m_{\rho }^{2}+\frac{5}{6}\right\} -\frac{z}{2}%
\right] \ .  \label{def: Z factors}
\end{eqnarray}
Adding all the contributions we obtain 
\begin{eqnarray}
\lefteqn{\Pi _{\rho L}(p;T=0)=}  \nonumber \\
&&\frac{N}{2}\frac{g^{2}}{(4\pi )^{2}}m_{\rho }^{2}\Biggl[ z\delta -\frac{37%
}{4}-\frac{1}{6\delta }-\frac{71}{24}\delta +\frac{11a^{2}}{72}\delta +\frac{%
\delta (1-\delta ^{2})}{12}\ln (-\delta )-\frac{a^{2}}{12}\delta \ln
(-\delta )  \nonumber \\
&&\quad +\frac{(4-\delta )^{2}(12+20\delta +\delta ^{2})}{6\sqrt{4-\delta }%
\sqrt{\delta }}\tan ^{-1}\sqrt{\frac{\delta }{4-\delta }}-\frac{(1-\delta
)^{3}(1+10\delta +\delta ^{2})}{6\delta ^{2}}\ln (1-\delta )\Biggr] \ ,
\end{eqnarray}
where $\delta $ is defined by  $\delta \equiv p^{2}/m_{\rho }^{2}$.
By requiring that $m_{\rho }$ becomes pole mass: 
\begin{equation}
\mbox{\rm Re}\, \Pi _{\rho L}(p^{2}=m_{\rho }^{2};T=0)=0\ ,
\end{equation}
this finite part $z$ is determined as 
\begin{equation}
z=\frac{99}{8}-\frac{11}{72}a^{2}-\frac{11\sqrt{3}\pi }{4}\ .
\label{z factor}
\end{equation}

The temperature dependence of the self-energy of $\rho $ meson obtained from
the thermal pion and $\rho $ meson is calculated by using the standard
imaginary time formalism.\cite{Matsubara} The temperature dependent parts, $%
\Delta \Pi (p_{0},\vec{p};T)\equiv \Pi (p_{0},\vec{p};T)-\Pi (p_{0},\vec{p}%
;T=0)$, are given by 
\begin{eqnarray}
\lefteqn{\mbox{\rm Re}\,
\Delta \Pi _{\rho L}^{(a)}(p_{0},\vec{p}=0;T)=\frac{N}{2}\frac{%
g^{2}}{\pi ^{2}}\frac{a^{2}}{12}G_{2}\ ,}  \nonumber \\
\lefteqn{\mbox{\rm Re}\, 
\Delta \Pi _{\rho L}^{(b)}(p_{0},\vec{p}=0;T)=\frac{N}{2}\frac{%
g^{2}}{\pi ^{2}}\frac{1}{12}G_{2}\ ,}  \nonumber \\
\lefteqn{\mbox{\rm Re}\, 
\Delta \Pi _{\rho L}^{(c)}(p_{0},\vec{p}=0;T)=}  \nonumber \\
&&\noindent \frac{N}{2}\frac{g^{2}}{\pi ^{2}}\Biggl[ \frac{-(4m_{\rho
}^{2}-p_{0}^{2})(m_{\rho }^{2}+p_{0}^{2})}{4m_{\rho }^{2}}F_{3}^{2}+\frac{%
-4m_{\rho }^{4}+9m_{\rho }^{2}p_{0}^{2}+p_{0}^{4}}{12m_{\rho }^{4}}F_{3}^{4}-%
\frac{1}{3m_{\rho }^{2}}F_{3}^{6}  \nonumber \\
&&\qquad +\frac{(m_{\rho }^{2}+p_{0}^{2})(m_{\rho }^{2}-p_{0}^{2})^{2}}{%
m_{\rho }^{2}}H_{1}^{2}+\frac{(m_{\rho }^{2}+p_{0}^{2})(m_{\rho
}^{2}-p_{0}^{2})^{2}}{3m_{\rho }^{4}}H_{1}^{4}  \nonumber \\
&&\qquad {}+\frac{p_{0}^{2}}{3m_{\rho }^{2}}\left\{ \frac{-(m_{\rho
}^{2}-p_{0}^{2})(11m_{\rho }^{2}+p_{0}^{2})}{m_{\rho }^{2}}K_{4}-4K_{6}+%
\frac{p_{0}^{2}}{4m_{\rho }^{2}}G_{2}\right\} \Biggr] \ ,  \nonumber \\
\lefteqn{\mbox{\rm Re}\,
\Delta \Pi _{\rho L}^{(d)}(p_{0},\vec{p}=0;T)=\frac{N}{2}\frac{%
g^{2}}{\pi ^{2}}\left[ -2J_{1}^{2}-\frac{1}{3m_{\rho }^{2}}\left(
I_{4}-J_{1}^{4}\right) \right] \ ,}  \nonumber \\
\lefteqn{\mbox{\rm Re}\,
\Delta \Pi _{\rho L}^{(e)}(p_{0},\vec{p}=0;T)=-\frac{N}{2}\frac{%
g^{2}}{\pi ^{2}}\frac{1}{6}G_{2}\ ,}  \label{func: T dep}
\end{eqnarray}
where functions $F$, $G$, $H$, $I$, $J$ and $K$ are defined in Appendix B.

Let us first consider the low temperature region $T\ll m_{\rho }$. The
functions $F$, $H$ and $J$ are suppressed by $e^{-m_{\rho }/T}$, and gives
negligible contribution.
Noting that in the low energy limit ($p_{0}\rightarrow 0$), 
$G_{2}\rightarrow I_{2}=(\pi^{2}/6)T^{2}$, 
we find
that the contribution to the $\rho $-meson propergator at low energy is
order $T^{2}$:
\begin{equation}
\Delta \Pi(p_0=0,\vec{p}=0;T) \approx \frac{N}{2} 
\frac{g^2}{72} (a^2-1) T^2 \ .
\end{equation}
Since $a\simeq2$, 
{\it the low energy mass parameter decreases as $T^2$}.
However, on mass-shell of $\rho $ meson, $p_{0}\approx
m_{\rho }$, $G_{2}$ term in (b), (c) and (e) are cancelled and 
\begin{equation}
\Delta \Pi (p_{0}=m_{\rho },\vec{p}=0;T)\approx 
\frac{N}{2}\frac{g^2}{\pi ^{2}} \frac{a^{2}}{12}G_{2}\ .
\end{equation}
Moreover, since $G_{2}\approx -\frac{\pi ^{4}}{15}\frac{T^{4}}{m_{\rho }^{2}}
$, {\it the }$\rho ${\it -meson mass increases as }$T^{4}${\it \ at low
temperature dominated by pion-loop effect.}

\begin{figure}[tbph]
\begin{center}
\ \epsfbox{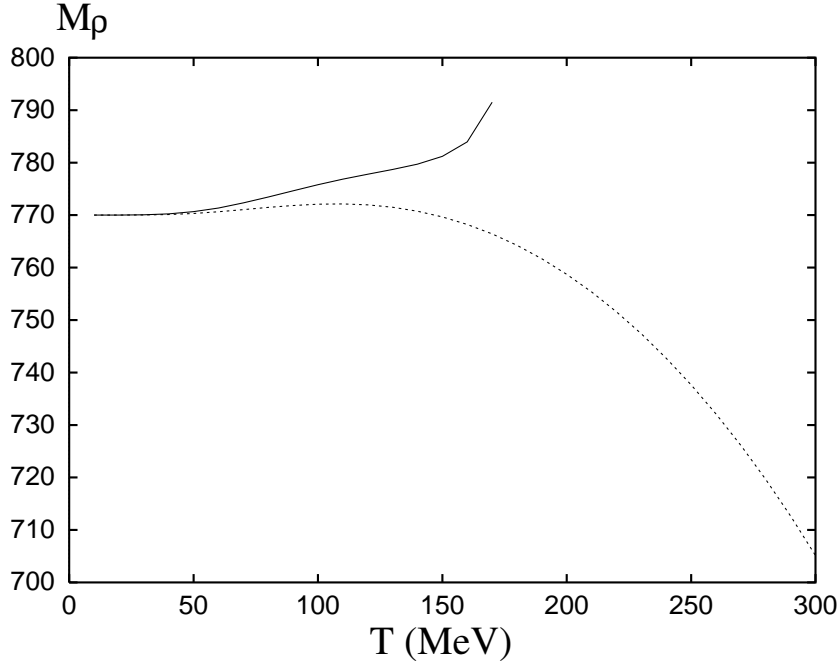}
\end{center}
\caption[]{ The temperature dependence of the pole mass of $\rho $
meson. The dotted line denotes the pole mass derived from the pion
loop contribution 
only, while the solid line denotes that from the total contribution.}
\label{pmassp}
\end{figure}
Next we study the temperature dependence of the pole mass at slightly higher
temperature by fully using the expression (\ref{func: T dep}). The pole
mass is defined as
\begin{equation}
M_{\rho }^{2}-m_{\rho }^{2}+
\mbox{\rm Re}\, \Pi _{\rho L}(p_{0}=M_{\rho },\vec{p}%
=0;T)=0\ .  \label{eq: pole mass}
\end{equation}
We show the temperature dependence of the pole mass by a solid line in Fig.~%
\ref{pmassp}. 
For $T>$ 170 MeV there is no solution of Eq.~(\ref{eq: pole mass}) below
1540 MeV when we include all the contributions. Then the solid line
terminates at $T\simeq 170$\thinspace MeV. 
Higher order loop effects may be needed to study such high temperature
region.
Until $T\simeq 170$\thinspace MeV
the mass of $\rho $ meson increases with temperature. The dotted line
denotes the pole mass when we include only the pion loop, where for
consistency we replace $z$ in Eq.~(\ref{z factor}) with 
\begin{equation}
z=-\frac{11}{72}a^{2}\ .  \label{z factor 2}
\end{equation}
Figure~\ref{pmassp} shows that in the high temperature region the pion-loop
correction decreases the $\rho $-meson mass, while the $\rho $ loop gives a
large positive correction, and as a result the $\rho $-meson mass increases
slightly with temperature.

\section{Summary and Discussion}

\label{sec: summary}In this paper we studied the temperature dependence of
the pion decay constant and the $\rho $-meson mass by including the thermal
pion and $\rho $-meson effects systematically in the hidden local symmetry
model. The thermal pion gives a dominant contribution to $f_{\pi }(T)$, and
the $\rho $ meson gives a small correction. The low temperature limit of our
result is consistent with the chiral perturbation prediction.\cite
{Gasser-Leutwyler:finiteT:1987} The inclusion of $\rho $ meson may be a good
approximation to ${\cal O}(p^{4})$ contribution to chiral perturbation
analysis.

We also studied the temperature dependence of the parameter $a$, which is
related to the successful phenomenological predictions of hidden local
symmetry. We showed that the parameter $a$ was stable as the temperature
increased, $a(T)\simeq2$. 
This may suggest that the 
phenomenological predictions of hidden local symmetry
holds at finite temperature in good
approximation. Our result shows that the parameter $a$ does not change at
all if we take $a=1$ from the beginning. 
This implies that in the ``vector
limit''\cite{Georgi:vectorlimit} 
the thermal effects of the $\rho$ meson do
not induce deviation from $a=1$.

We calculated the temperature dependence of 
the mass of $\rho$ meson.
We showed that the low-energy mass parameter decreased as $T^2$
in the low temperature region.
However, the $\rho $-meson pole mass 
increases as $T^{4}$
dominated by pion-loop effect, which is consistent with the result by the
current algebra analysis.\cite{Dey-Eletsky-Ioffe} In the high temperature
region pion gives a negative contribution, while the $\rho $ loop gives a
large positive correction, and 
the $\rho $-meson mass 
increases with
temperature. The correction from pion loop in our model does not seem to
agree with previous analysis\cite{rhomass:pi} at slightly higher
temperature. The difference between the result of our model and that of
previous models comes from the fact that there is no pion tad-pole
contribution at one loop 
in our model.
The tad-pole diagram gives
a relatively large
positive contribution to the $\rho $-meson mass in the previous models.
Instead in our model, $\rho $ meson gives a positive contribution, and it
overcomes the pion-loop contribution.
Although each correction to the $\rho $%
-meson mass is small, it is interesting to point out that the $\rho $-loop
contribution is bigger than 
the pion-loop contribution
for $T\geq 50$\thinspace
MeV. Our analysis implies that the inclusion of $\rho $ meson itself is
important to study the temperature dependence of $\rho $-meson mass near
critical temperature.

In the high temperature region one-loop approximation might be too crude to
study the temperature dependences. Although we can expect that the
qualitative structure does not change, it seems to be interesting to include
the higher order effect, for example, by using the temperature dependent
renormalization group analysis\cite{Matsumoto-Nakano-Umezawa}.

Finally we make a comment on the correction to the $\rho$-meson mass
from 
the $\omega$-$\pi$ loop. In this paper we did not include the anomalous
interaction terms  such as $\omega$-$\rho$-$\pi$, 
since they are ${\cal O}(p^4)$ terms in a chiral counting rule~[17]
and expected to give higher order effects.
The $\omega$-$\pi$ loop correction to the $\rho$-meson self-energy 
is ${\cal O}(p^8)$, while the $\rho$-meson loop generates 
${\cal O}(p^4)$ correction.
Apparently the $\omega$-$\pi$ loop does not generate any correction
to the 
low-energy mass parameter. 
To the pole mass, there is a suppression factor 
$\left(g/4\pi\right)^4$ for the $\omega$-$\pi$ loop correction 
compared with
the $\rho$-loop correction, then we expect that the correction is
small.

\section*{Acknowledgment}

We would like to thank to Prof. Koichi Yamawaki for suggesting this subject
and for stimulating discussions.
We would like to thank also to people
in Nagoya University where
the final stage of this work was done.
We are also grateful to Prof. Joe Schechter for useful comment.

\newpage \appendix

\begin{center}
{\Large {\sc Appendices }}
\end{center}

\section{Polarization Tensors}

\label{app: polar}

At finite temperature, the polarization tensor is no longer restricted to be
Lorentz covariant, but only $O(3)$ covariant. Then the polarization tensors
can be expressed by four independent symmetric $O(3)$ tensors. Here we list
up the polarization tensors at finite temperature:\cite{Toimele,NDoray} 
\begin{eqnarray}
P_{T\mu\nu} &=&
  g_{\mu i} \left( 
      \delta_{ij} - \frac{\vec{p}_i\vec{p}_j}{\left\vert\vec{p}\right\vert^2}
  \right)g_{j \nu} \nonumber \\
&=& \left\{
\begin{array}{l}
P_{T00} = P_{T0i} = P_{Ti0} = 0 \ , \\ 
P_{Tij} = \delta_{ij} - \frac{\vec{p}_i\vec{p}_j}{\left\vert\vec{p}%
\right\vert^2} \ ,
\end{array}
\right.
\nonumber \\
P_{L\mu\nu} &\equiv& - \left( g_{\mu\nu} - \frac{p_\mu p_\nu}{p^2} \right) -
P_{T\mu\nu}   \nonumber \\
&=&
\left(g_{\mu 0}-\frac{p_\mu p_0}{p^2} \right) 
\frac{p^2}{\left\vert\vec{p}\right\vert^2}
\left(g_{0 \nu}-\frac{p_0 p_\nu}{p^2} \right) 
\ , \nonumber \\
P_{C\mu\nu} &\equiv& \frac{1}{\sqrt{2}\left\vert\vec{p}\right\vert} \left[
\left( g_{\mu0} - \frac{p_\mu p_0}{p^2} \right) p_\nu + p_\mu \left(
g_{0\nu} - \frac{p_0 p_\nu}{p^2} \right) \right] \ ,  \nonumber \\
P_{D\mu\nu} &\equiv& \frac{p_\mu p_\nu}{p^2} \ ,  \label{polar tensor}
\end{eqnarray}
where $p^\mu = (p_0,\vec{p})$ is four-momentum.

The following formulae are convenient:
\begin{eqnarray}
P_{L\mu\alpha} P_{L}^{\alpha\nu} &=& - {P_{L\mu}}^\nu \ ,
\nonumber\\
P_{T\mu\alpha} P_{T}^{\alpha\nu} &=& - {P_{T\mu}}^\nu \ ,
\nonumber\\
P_{C\mu\alpha} P_{C}^{\alpha\nu} &=& \frac{1}{2}
\left( {P_{L\mu}}^\nu + {P_{D\mu}}^\nu \right) \ ,
\nonumber\\
P_{D\mu\alpha} P_{D}^{\alpha\nu} &=& {P_{D\mu}}^\nu \ ,
\nonumber\\
P_{L\mu\alpha} P_{T}^{\alpha\nu} &=& 
P_{C\mu\alpha} P_{T}^{\alpha\nu} =
P_{D\mu\alpha} P_{T}^{\alpha\nu} =
P_{D\mu\alpha} P_{L}^{\alpha\nu} = 0 \ ,
\nonumber\\
P_{C\mu\alpha} P_{L}^{\alpha\nu} &=& 
- P_{D\mu\alpha} P_{C}^{\alpha\nu} =
- \frac{p_\mu}{\sqrt{2}\left\vert\vec{p}\right\vert}
\left( g_{0}^\nu - \frac{p_0 p^\nu}{p^2} \right) \ .
\end{eqnarray}

For the transversal tensor $p_\mu \Pi^{\mu\nu}(p)=0$ we can 
decompose it into
\begin{equation}
  \Pi_{\mu\nu}(p) = P_{L\mu\nu}\Pi_{L}(p)+ P_{T\mu\nu}\Pi_{T}(p) \ ,
\end{equation}
where $\Pi_{L}$ and $\Pi_T$ are given by
\begin{eqnarray}
   \Pi_{L} &=& \frac{p^2}{\left\vert\vec{p}\right\vert^2}\Pi_{00} \ ,
    \nonumber \\
   \Pi_{T} &=& -\frac{1}{2}
      \left\{ \Pi^{j}_{j} 
    + \frac{p_0^2}{\left\vert\vec{p}\right\vert^2}\Pi_{00}
   \right\} \ .
\end{eqnarray}

\section{Functions}

\label{app: functions}

Here we list the functions used in this paper. 

Functions used in the expressions of $f_\pi$ and $f_\sigma$ 
in Eqs.~(\ref{eq: fpi}) and (\ref{fsigma: T})
are defined as follows;
\begin{eqnarray}
I_{n}(T) &\equiv &\int_{0}^{\infty }d{\rm k}
\frac{{\rm k}^{n-1}}{e^{{\rm k}/T}-1}
=\widetilde{I}_{n}T^{n} \ ,  \nonumber \\
&&\widetilde{I}_{n}=
\int_{0}^{\infty }d{y}\frac{y^{n-1}}{e^y-1}
=(n-1)! \, \zeta(n)\ ,\nonumber \\
&&\widetilde{I}_{2}=\frac{\pi ^{2}}{6}\ ,\quad
\widetilde{I}_{4}=\frac{\pi ^{4}}{15}\ ,\quad
\widetilde{I}_{6}=\frac{%
8\pi ^{6}}{63}\ ,   \nonumber \\
J_{m}^{n}(m_{\rho };T) &\equiv &\int_{0}^{\infty }d{\rm k}\frac{1}{e^{\omega
/T}-1}\frac{{\rm k}^{n}}{\omega ^{m}}\qquad ;\quad
n,m:\,\mbox{\rm integer}\ ,  \nonumber \\
&&\qquad \omega \equiv \sqrt{{\rm k}^{2}+m_{\rho }^{2}}\ .
\label{function 1}
\end{eqnarray}
We also define 
the functions in the $\rho$-meson propagator as follows:
\begin{eqnarray}
F_{3}^{n}(p_{0};m_{\rho };T) &\equiv &\int_{0}^{\infty }d{\rm k}{\cal P}%
\frac{1}{e^{\omega /T}-1}
\frac{4{\rm k}^{n}}{\omega (4\omega ^{2}-p_{0}^{2})} \ ,  \nonumber \\
G_{n}(p_{0};T) &\equiv &\int_{0}^{\infty }d{\rm k}{\cal P}
\frac{{\rm k}^{n-1}%
}{e^{{\rm k}/T}-1}\frac{4{\rm k}^{2}}{4{\rm k}^{2}-p_{0}^{2}}  \nonumber \\
&=& I_n(T) +
 \int_{0}^{\infty }d{\rm k}{\cal P}\frac{{\rm k}^{n-1}%
}{e^{{\rm k}/T}-1}\frac{p_{0}^{2} }{4{\rm k}^{2}-p_{0}^{2}} \,  \nonumber 
\\
H_{1}^{n}(p_{0};m_{\rho };T) &\equiv &\int_{0}^{\infty }d{\rm k}{\cal P}%
\frac{1}{e^{\omega /T}-1}\frac{{\rm k}^{n}}{\omega }\frac{1}{(m_{\rho
}^{2}-p_{0}^{2})^{2}-4{\rm k}^{2}p_{0}^{2}}\ ,  \nonumber \\
K_{n}(p_{0};m_{\rho };T) &\equiv &\int_{0}^{\infty }d{\rm k}{\cal P}\frac{%
{\rm k}^{n-1}}{e^{{\rm k}/T}-1}
\frac{1}{(m_{\rho }^{2}-p_{0}^{2})^{2}-4{\rm k%
}^{2}p_{0}^{2}}\ ,  \label{function 2}
\end{eqnarray}
where ${\cal P}$ denotes the principal part.

\section{$\rho$ and $\sigma$ propagators}

\label{app: rho sigma propagator}

\begin{figure}[htbp]
\begin{center}
\ \epsfbox{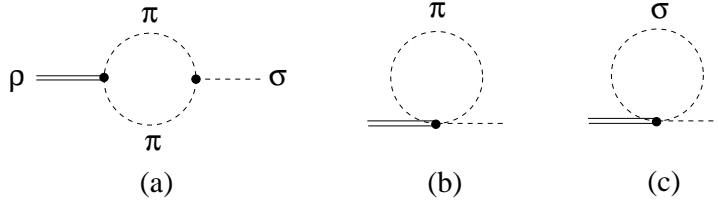}
\end{center}
\caption[]{ Diagrams contributing to the $\rho$-$\sigma$ mixing at
one loop in the Landau gauge.}
\label{fig: rho sigma}
\end{figure}

We introduced an $R_\xi$-like gauge-fixing term (\ref{Lag: GF FP}) for
eliminating the tree-level $\rho$-$\sigma$ mixing existing in the Lagrangian
(\ref{Lag}). Generally, a new $\rho$-$\sigma$ mixing is generated by
one-loop effects. At zero temperature one-loop corrections do not generate
any corrections to the $\rho$-$\sigma$ mixing when we use the Landau gauge.
Here we calculate the $\rho$-$\sigma$ mixing at finite temperature. There
are three diagrams contributing to the $\rho$-$\sigma$ mixing at one loop in
the Landau gauge: (a) $\pi$-loop; (b) $\pi$ tad-pole; (c) $\sigma$ tad-pole,
which are shown in Fig.~\ref{fig: rho sigma}. 
These diagrams give corrections to the $\rho$-$\sigma$ mixing given by 
\begin{equation}
F_\mu(p_0,\vec{p};T) = \left[ \left( N(a-1) + \frac{1}{N} \right) \frac{%
(a+1) g}{8\pi^2 f_\sigma} I_2 \right] p_\mu \ ,  \label{eq: rho sigma}
\end{equation}
where $I_2$ is defined in Eq.~(\ref{function 1}). It is remarkable to see
that all the contributions are proportional to four-momentum $p_\mu$.

When $\rho$-$\sigma$ mixing is proportional to four-momentum, the
inverse-propagator matrix for $\rho$ and $\sigma$, in general, takes the
form given by 
\begin{equation}
\Delta^{-1}_{(\mu,\nu)} = \left( 
\begin{array}{cc}
\widetilde{D}_\sigma^{-1} & i F p_\nu \\ 
- i F p_\mu & \widetilde{D}^{-1}_{\mu\nu}
\end{array}
\right) \ .
\end{equation}
Here $\widetilde{D}_\sigma^{-1}$ and $\widetilde{D}^{-1}_{\mu\nu}$ are
inverse $\sigma$ and $\rho$ propagators before diagonalization. 
The inverse $%
\rho$ propagator $\widetilde{D}_{\mu\nu}^{-1}$ is projected to four
polarizations by the polarization tensor given in Eq.~(\ref{polar tensor}): 
\begin{equation}
\widetilde{D}_{\mu\nu}^{-1} \equiv P_{T\mu\nu} \widetilde{\Pi}_T +
P_{L\mu\nu} \widetilde{\Pi}_L + P_{C\mu\nu} \widetilde{\Pi}_C + P_{D\mu\nu} 
\widetilde{\Pi}_D \ .  \label{rho prop inv}
\end{equation}
After some calculations we obtain the $\rho$-$\sigma$ propagator matrix: 
\begin{equation}
\Delta^{-1}_{(\mu,\nu)} = \left( 
\begin{array}{cc}
A & 
\begin{array}{cc}
ip_0 B & i p_j C
\end{array}
\\ 
\begin{array}{c}
- ip_0 B \\ 
- i p_i C
\end{array}
& D_{\mu\nu}
\end{array}
\right) \ ,
\end{equation}
where 
\begin{eqnarray}
A &=& \frac{\widetilde{D}_\sigma \left[ \widetilde{\Pi}_L \widetilde{\Pi}_D
- (\widetilde{\Pi}_C)^2/2 \right] }{\widetilde{\Pi}_L \widetilde{\Pi}_D - (%
\widetilde{\Pi}_C)^2/2 - p^2 F^2 \widetilde{D}_\sigma \widetilde{\Pi}_L } \ ,
\nonumber \\
B &=& \frac{\widetilde{D}_\sigma F \left[ \overline{p} \widetilde{\Pi}_C /%
\sqrt{2} - p_0 \widetilde{\Pi}_L \right] }{\overline{p} \left[ \widetilde{\Pi%
}_L \widetilde{\Pi}_D - (\widetilde{\Pi}_C)^2/2 - p^2 F^2 \widetilde{D}%
_\sigma \widetilde{\Pi}_L \right] } \ ,  \nonumber \\
C &=& \frac{\widetilde{D}_\sigma F \left[ p_0 \widetilde{\Pi}_C /\sqrt{2} - 
\overline{p} \widetilde{\Pi}_L \right] }{ p_0 \left[ \widetilde{\Pi}_L 
\widetilde{\Pi}_D - (\widetilde{\Pi}_C)^2/2 - p^2 F^2 \widetilde{D}_\sigma 
\widetilde{\Pi}_L \right] } \ ,
\end{eqnarray}
and 
\begin{equation}
D_{\mu\nu} = \frac{P_{T\mu\nu}}{\widetilde{\Pi}_T} + \frac{ P_{L\mu\nu}
\left[ \widetilde{\Pi}_D - p^2 \widetilde{D}_\sigma F^2 \right] +
P_{C\mu\nu} \widetilde{\Pi}_C + P_{D\mu\nu} \widetilde{\Pi}_L }{\widetilde{%
\Pi}_L \widetilde{\Pi}_D - (\widetilde{\Pi}_C)^2/2 - p^2 F^2 \widetilde{D}%
_\sigma \widetilde{\Pi}_L } \ .
\end{equation}

In the $R_\xi$ gauge, the inverse free propagator of $\rho$ meson is give by 
\begin{equation}
{\ D_0^{-1} }_{\mu\nu} = - \left( P_{L\mu\nu} + P_{T\mu\nu} \right)
(p^2-m_\rho^2) + \frac{p^2-\alpha m_\rho^2}{\alpha} \frac{p_\mu p_\nu}{p^2}
\ ,
\end{equation}
where $\alpha$ is the gauge parameter.
The inverse full propagator is defined by $D^{-1}_{\mu\nu} = {D_0^{-1}}%
_{\mu\nu} - \Pi_{\mu\nu}$, where the $\rho$ meson self-energy $%
\Pi^{\mu\nu}_{\rho}$ is expanded by four independent polarization tensors
listed in Eq.~(\ref{polar tensor}): 
\begin{equation}
\Pi^{\mu\nu}_{\rho} = P_T^{\mu\nu} \Pi_{\rho T} + P_L^{\mu\nu} \Pi_{\rho L}
+ P_C^{\mu\nu} \Pi_{\rho C} + P_D^{\mu\nu} \Pi_{\rho D} \ .
\label{Pi: decomp}
\end{equation}
By using these quantities, four components of the inverse $\rho$ propagator
in Eq.~(\ref{rho prop inv}) are given by 
\begin{eqnarray}
\widetilde{\Pi}_T &=& - \left( p^2 - m_\rho^2 + \Pi_{\rho T} \right) \ , 
\nonumber \\
\widetilde{\Pi}_L &=& - \left( p^2 - m_\rho^2 + \Pi_{\rho L} \right) \ , 
\nonumber \\
\widetilde{\Pi}_C &=& - \Pi_{\rho C} \ ,  \nonumber \\
\widetilde{\Pi}_D &=& \frac{p^2-\alpha m_\rho^2}{\alpha} - \Pi_{\rho D} \ .
\end{eqnarray}
If we take the Landau gauge $\alpha=0$, the $\rho$-$\sigma$ propagator
matrix reduces to a simple form: 
\begin{eqnarray}
&& A = \widetilde{D}_\sigma \ , \quad B = C = 0 \ ,  \nonumber \\
&& D_{\mu\nu} = - \frac{P_{T\mu\nu}}{p^2-m_\rho^2+\Pi_{\rho T}} - \frac{%
P_{L\mu\nu}}{p^2-m_\rho^2+\Pi_{\rho L}} \ .  \label{rho propagator 2}
\end{eqnarray}

\end{document}